\documentclass{article}

\usepackage{PRIMEarxiv}

\usepackage[utf8]{inputenc} 
\usepackage[T1]{fontenc}    
\usepackage{hyperref}       
\usepackage{url}            
\usepackage{booktabs}       
\usepackage{amsfonts}       
\usepackage{nicefrac}       
\usepackage{microtype}      
\usepackage{lipsum}
\usepackage{fancyhdr}       
\usepackage{graphicx}       
\graphicspath{{media/}}     
\usepackage{amssymb,amsmath}
\usepackage{graphicx}
\usepackage{hyperref}
\usepackage{listings}
\usepackage{xcolor}
\usepackage{longtable} 
\usepackage{graphicx} 
\usepackage[backend=bibtex,style=numeric]{biblatex}
\usepackage{geometry}
\usepackage{tabularx} 
\usepackage{tabu}
\usepackage{adjustbox}

\definecolor{codegreen}{rgb}{0,0.6,0}
\definecolor{codegray}{rgb}{0.5,0.5,0.5}
\definecolor{codepurple}{rgb}{0.58,0,0.82}
\definecolor{backcolour}{rgb}{0.95,0.95,0.92}

\lstdefinestyle{mystyle}{
    backgroundcolor=\color{backcolour},   
    commentstyle=\color{codegreen},
    keywordstyle=\color{magenta},
    numberstyle=\tiny\color{codegray},
    stringstyle=\color{codepurple},
    basicstyle=\ttfamily\footnotesize,
    breakatwhitespace=false,         
    breaklines=true,                 
    captionpos=b,                    
    keepspaces=true,                 
    numbers=left,                    
    numbersep=5pt,                  
    showspaces=false,                
    showstringspaces=false,
    showtabs=false,                  
    tabsize=2
}
\title{RMCDA: The Comprehensive R Library for Applying Multi-Criteria Decision Analysis Methods

}

\author{
  Annice Najafi$^{1}$, Shokoufeh Mirzaei$^{1,*}$ \\
  $^{1}$California Polytechnic Institute of Technology, Pomona \\
  Pomona, CA\\
  $^{*}$Corresponding author \\
  \texttt{smirzaei@cpp.edu}
}

\begin{document}
\maketitle

\begin{abstract}
Multi-Criteria Decision Making (MCDM) is a branch of operations research used in a variety of domains from health care to engineering to facilitate decision-making among multiple options based on specific criteria. Several \texttt{R} packages have been developed for the application of traditional MCDM approaches. However, as the discipline has advanced, many new approaches have emerged, necessitating the development of innovative and comprehensive tools to enhance the accessibility of these methodologies. Here, we introduce \texttt{RMCDA}, a comprehensive and universal \texttt{R} package that offers access to a variety of established MCDM approaches (e.g., \texttt{AHP}, \texttt{TOPSIS}, \texttt{PROMETHEE}, and \texttt{VIKOR}), along with newer techniques such as Stratified MCDM (\texttt{SMCDM}) and the Stratified Best-Worst Method (\texttt{SBWM}). Our open source software intends to broaden the practical use of these methods through supplementary visualization tools and straightforward installation. 
\end{abstract}

\keywords{MCDA \and MCDM \and \texttt{R} package \and Universal Library \and Decision Making}

\section*{Metadata}

\begin{table}[!h]
\caption{Code metadata}
\vspace{5mm}
\begin{tabular}{|l|p{6.5cm}|p{6.5cm}|}
\hline
\textbf{Nr.} & \textbf{Code metadata description} & \textbf{Metadata} \\
\hline
C1 & Current code version & 0.3 \\
\hline
C2 & Permanent link to code/repository & \url{https://github.com/AnniceNajafi/RMCDA} \\
\hline
C3 & Permanent link to Reproducible Capsule & None\\
\hline
C4 & Legal Code License   & MIT License \\
\hline
C5 & Code versioning system used & git \\
\hline
C6 & Software code languages, tools and services used & \texttt{R} \\
\hline
C7 & Compilation requirements, operating environments \& dependencies & \texttt{R} package: \texttt{dplyr},
    \texttt{stats},
    \texttt{igraph},
    \texttt{fmsb},
    \texttt{lpSolve},
    \texttt{MASS},
    \texttt{matlib},
    \texttt{nloptr},
    \texttt{matrixStats},
    \texttt{pracma},.\\
\hline
C8 & Link to developer documentation/manual & \url{https://github.com/AnniceNajafi/RMCDA/blob/main/RMCDA_0.3.pdf} \\
\hline
C9 & Support email for questions & \texttt{annicenajafi27@gmail.com}\\
\hline
\end{tabular}
\label{codeMetadata} 
\end{table}

\section{Motivation and significance}
Multi-Criteria Decision Analysis (MCDA) methods have been historically used to compare and rank options based on multiple criteria \cite{mardani2015multiple, zionts1979mcdm}. Over the years, various approaches, such as the Analytic Hierarchy Process (\texttt{AHP}) \cite{saaty1980analytic} and the Technique for Order of Preference by Similarity to Ideal Solution (\texttt{TOPSIS}) \cite{tzeng2011multiple}, have been extensively developed. Bibliographic analyses reveal that \texttt{TOPSIS}, \texttt{AHP}, \texttt{VIKOR} (VIsekriterijumsko KOmpromisno Rangiranje), \texttt{PROMETHEE} (Preference Ranking Organization Method for Enrichment Evaluation), and \texttt{ANP} (Analytic Network Process) are among the most popular MCDA methods \cite{rocha2024review}.

Numerous \texttt{R} packages (e.g., \texttt{topsis} \cite{topsisR}, \texttt{ahpsurvey} \cite{ahpsurveyR}, \texttt{MCDA} \cite{MCDAR, bigaret2017supporting}, \texttt{AHPtools} \cite{AHPtools}, \texttt{PROMETHEE} \cite{ishizaka2018visual, greco2019methodological}, and others) have simplified the application of these established techniques. However, the scope of their functionality is limited to a few methods. Nonetheless, newer methods are not widely available in \texttt{R}, including the Stratified MCDM (\texttt{SMCDM}) \cite{asadabadi2018stratified}, and Stratified Best-Worst Method (\texttt{SBWM}) \cite{torkayesh2021sustainable}. Despite the recent introduction of these methods, they have been classified as having above-average Category Normalized Citation Impact by Web Of Science, highlighting their growing significance in MCDA research. These methods handle more complex and “stratified” problems, such as decisions that must account for changing probabilities or uncertain future states, such as policy planning under shifting economic or climate conditions.

While implementations of traditional MCDA methods exist in other programming environments (e.g., \texttt{pyDecision} \cite{pereira2024enhancing}, \texttt{pymcdm} \cite{kizielewicz2023pymcdm}, \texttt{pyrepo} \cite{wkatrobski2022pyrepo}, \texttt{JMcDM} \cite{satman2021jmcdm}, and even standalone tools such as DIVIZ \cite{meyer2012diviz}), there is currently no equivalent implementation of stratified MCDM approaches in these languages. Moreover, even though the \texttt{R} ecosystem has a large user base among researchers and data analysts, no package provides the same extent of MCDA methods available in other programming languages.

Our goal is therefore to offer a comprehensive \texttt{R} package (\texttt{RMCDA}) encompassing both new and established methods. By filling this gap, \texttt{RMCDA} can potentially broaden the accessibility of these methods, enabling both deterministic and probabilistic forms of decision-making. 

\section{Comparison with existing \texttt{R} libraries and other tools}
We used the R package `pkgsearch'\cite{pkgsearch} to find R packages with the keywords `MCDA' and `MCDM'. A total of 15 packages were found. The ones associated with the keyword `MCDA' were `cocosoR', `MCDA', `smaa', `PROMETHEE', `RXMCDA', `brisk', and `ConsRank'. Packages associated with the `MCDM' keyword were `ahptopsis2n', `topsis', `FuzzyMCDM', `sapevom', `IFMCDM', `fucom', `rafsi', `SDEFSR'. Of these packages, the `MCDA' and `MCDM' packages were the most general and covered the most MCDA methods (we use the term MCDA and MCDM interchangeably throughout this manuscript). 

\begin{figure}
    \centering
    \includegraphics[width=0.8\linewidth]{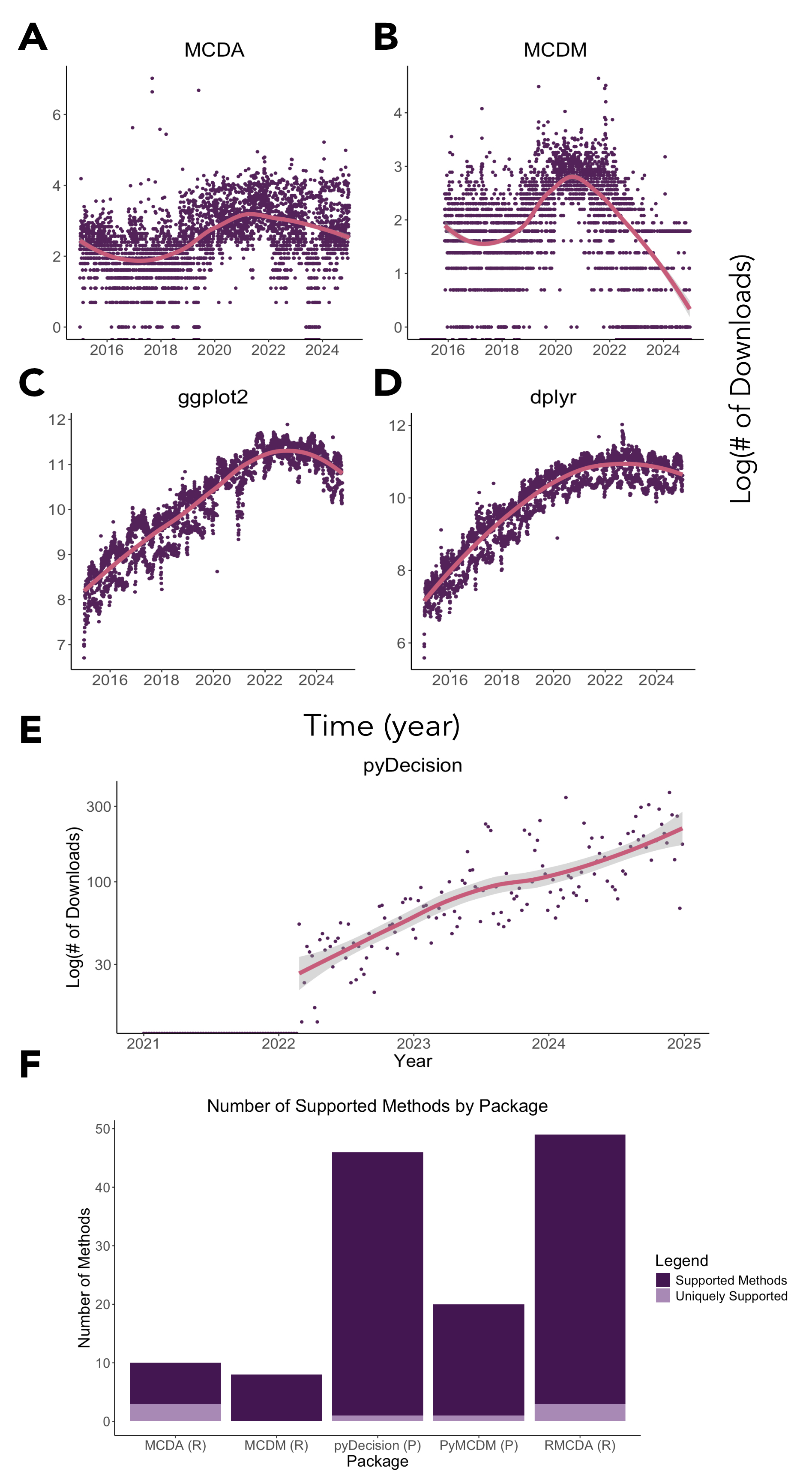}
    \caption{\textbf{A-E show the number of downloads over time by package. F compares the number of supported methods by package.}}
    \label{cranlogs}
\end{figure}

Next, we used the `cranlogs' package in R \cite{cranlogs} to find the number of downloads of the `MCDA' and `MCDM' packages over time and through Locally Estimated Scatterplot Smoothing (LOES) showed that both packages followed similar temporal patterns in terms of their number of downloads. As shown in Figure \ref{cranlogs}A-B, the temporal patterns of the two packages can be divided into three phases. First, the packages saw a decline in popularity between years 2016 and 2018, followed by an increase in the number of downloads from 2018 to 2021 eventually accompanied by a sudden decline in popularity. To understand whether this pattern is specific to the two MCDA packages in R or indirectly caused by the changing popularity of R as a programming language, we applied the same methodology to two of the most used R packages, `ggplot2', and `dplyr' and compared the trajectories. The results showed a monotonic increase in the number of downloads of both packages followed by a decline after 2022 (Figure \ref{cranlogs}C-D). This was inconsistent with the temporal patterns of the `MCDA' and `MCDM' packages. We also note that the number of downloads of these two packages over time is inconsistent with the number of publications related to MCDA \cite{srivastava2024multi} suggesting a decline in the popularity of these R packages among users despite the increase in the popularity of MCDA methods in general.

To understand if this pattern is specific to MCDA-related packages in R or consistent across packages in other programming languages, we used the `pypinfo' package in Python \cite{levwilk} to retrieve the number of downloads for one of the comprehensive Python packages for MCDA, the \texttt{pyDecision} package \cite{pereira2024enhancing}. The results showed a linear increase in the number of downloads of this package since its introduction, suggesting that the decline in the popularity of the most comprehensive R packages may be due to their lack of coverage of the newer MCDA methods (Figure \ref{cranlogs}E). To put this into perspective, we found the total number of MCDA methods included in each package and have plotted the results in Figure \ref{cranlogs}F. As demonstrated in the figure, the MCDA-related packages in Python offer significantly more methods than the previously introduced R packages. The figure also demonstrates that the MCDA package in R provides a significant number of unique methods that are not covered in other MCDA-related libraries, indicating that the package may include methods that are unpopular. 
Compared to other options, our RMCDA package stands out by offering widely used methods among researchers. For example, it includes COPRAS \cite{zavadskas2007multi}, SMCDM \cite{asadabadi2018stratified}, and SBWM \cite{torkayesh2021sustainable} that were previously unavailable in any R package. These methodologies have demonstrated strong academic impact, with above-average Category Normalized Citation Impact ratings on the Web of Science.

The RMCDA R package covers more than 50 popular MCDA and weighting methods; Analytical Hierarchy Process (AHP) \cite{saaty2004decision}, Analytical Network Process (ANP) \cite{saaty2006decision}, Additive Ratio ASsessment (ARAS) \cite{zavadskas2010new}, Borda \cite{borda1781m}, Best Worst Method (BWM) \cite{rezaei2015best}, Criterion Impact LOSs (CILOS) \cite{zavadskas2016integrated}, Combined Compromise Solution (CoCoSo) \cite{yazdani2019combined}, Combinative Distance-based ASsessment (CODAS) \cite{keshavarz2016new}, Copeland \cite{saari1996copeland}, Complex Proportional ASsessment (COPRAS) \cite{zavadskas2007multi}, Compromise RAnking and Distance from Ideal Solution (CRADIS) \cite{puvska2022evaluation}, CRiteria Importance Through Intercriteria Correlation (CRITIC) \cite{diakoulaki1995determining}, DEcision-MAking Trial and Evaluation Laboratory (DEMATEL) \cite{si2018dematel}, Evaluation based on Distance from Average Solution (EDAS) \cite{keshavarz2015multi}, ELimination Et Choix Traduisant la REalité (ELECTRE) \cite{roy1968classement}, Entropy \cite{shannon1948mathematical}, Grey Relational Analysis (GRA) \cite{kuo2008use}, Integrated Determination of Objective CRIteria Weights (IDOCRIW) \cite{zavadskas2016integrated}, Multi-Attributive Border Approximation Area Comparison (MABAC) \cite{pamuvcar2015selection}, Measuring Attractiveness by a Categorical Based Evaluation TecHnique (MACBETH) \cite{marcelino2019development}, Multi-Attribute Ideal Real Comparative Analysis (MAIRCA) \cite{hadian2022multi}, Multi-Attribute Ranking Approach (MARA) \cite{gligoric2022novel}, Measurement of Alternatives and Ranking based on COmpromise Solution (MARCOS) \cite{stevic2020sustainable}, Multi-Attribute Utility Theory \cite{keeney1993decisions}, Multi-Objective Optimization by Ratio Analysis (MOORA) \cite{brauers2006moora}, Multi-Objective Optimization on the basis of Simple Ratio Analysis (MOOSRA) \cite{jagadish2014green}, Multi-Objective Optimization on the basis of a Ratio Analysis plus the full MULTIplicative form (Multi-MOORA) \cite{brauers2006moora}, Outranking Compromise Ranking Approach (OCRA) \cite{madic2015selection}, Ordered Performance Analysis (OPA) \cite{ataei2020ordinal}, Organisation, Rangement Et Synthèse De Données Relatives À L’évaluation (ORESTE) \cite{roubens1982preference}, Position Index Value (PIV) \cite{mufazzal2018new}, Preference Ranking Organization Method for Enrichment Evaluation (PROMETHEE) \cite{brans2016promethee}, Preference Selection Index (PSI) \cite{maniya2010selection}, Ranking Alternatives using Full Subset Inference (RAFSI) \cite{vzivzovic2020eliminating}, Regime \cite{hinloopen1990qualitative}, Ranking Index Method (RIM) \cite{cables2016rim}, Range of Value (ROV) \cite{madic2016application}, Simple Additive Weighting (SAW) \cite{panjaitan2019simple}, Stratified Best Worst Method (SBWM) \cite{torkayesh2021sustainable}, Simple Evaluation of Complex Alternatives (SECA) \cite{keshavarz2018simultaneous}, Simple Multi-Attribute Rating Technique (SMART) \cite{olson1997decision}, Stratified Multi-Criteria Decision Making (SMCDM) \cite{asadabadi2018stratified}, Stable Preference Ordering Towards Ideal Solution (SPOTIS) \cite{dezert2020spotis}, Simple Ranking Method using reference Profiles (SRMP) \cite{khannoussi2022simple}, TOmada de Decisão Interativa e Multicritério (TODIM) \cite{gomes2009application}, Technique for Order of Preference by Similarity to Ideal Solution (TOPSIS) \cite{hwang1981methods}, VIšekriterijumsko KOmpromisno Rangiranje (VIKOR) \cite{opricovic2004compromise}, Weighted Aggregated Sum Product ASsessment (WASPAS) \cite{zavadskas2012optimization}, Weighted Product Method (WPM), Weighted Sum Method (WSM) \cite{san2012weighted}, Weighted Influence Nonlinear Gauge System (WINGS) \cite{michnik2013weighted}, and Weighted Influence Score Preference (WISP) \cite{stanujkic2021integrated}. 
Among the four MCDA packages that we investigated, only four methods are not included in the RCMDA package; Simulated Uncertainty Range Evaluations (SURE) \cite{hodgett2019sure}, Multi-Attribute Range Evaluations \cite{hodgett2014handling}, Utilité Additive (UTA) \cite{jacquet1982assessing}, and Characteristic Objects METhod (COMET) \cite{salabun2015characteristic}.
Table~\ref{comparisonTable} illustrates a short comparison of the key MCDA-related packages in different programming languages. By introducing RMCDA, we aim to provide a universal and comprehensive R library that enhances the accessibility to MCDA and criteria-weighing methods. We note that in the table we have not included the implementation of fuzzy variations of the methods (RMCDA only offers the fuzzy variation of the AHP method \cite{pehlivan2017comparison}).

\clearpage 

\begin{longtabu}[c]{|c|c|c|c|c|c|}
\caption{Comparison of select MCDA packages with \texttt{RMCDA}.}
\label{comparisonTable} \\
\hline
\textbf{Feature} & \texttt{MCDA(R)} & \texttt{MCDM(R)} & \texttt{PyMCDM(Python)} & \texttt{pyDecision(Python)} & \texttt{RMCDA(R)} \\
\hline
\endfirsthead

\hline
\textbf{Feature} & \texttt{MCDA(R)} & \texttt{MCDM(R)} & \texttt{PyMCDM(Python)} & \texttt{pyDecision(Python)} & \texttt{RMCDA(R)} \\
\hline
\endhead

\hline
\multicolumn{6}{r}{\textit{Continued on next page}} \\
\hline
\endfoot

\hline
\endlastfoot

\texttt{AHP}      & \checkmark & -          & -          & \checkmark & \checkmark \\
\texttt{ANP}      & - & -          & -          & - & \checkmark \\
\texttt{ARAS}      & - & -          & \checkmark          & \checkmark & \checkmark \\
\texttt{BORDA}      & - & -          & -          & \checkmark & \checkmark \\
\texttt{BWM}       & - & -          & - & \checkmark          & \checkmark \\
\texttt{CILOS}    & -          & - & \checkmark          & \checkmark          & \checkmark \\
\texttt{COCOSO}    & -          & - & \checkmark          & \checkmark          & \checkmark \\
\texttt{CODAS}          & -          & -          & \checkmark          & \checkmark          & \checkmark \\
\texttt{COMET}   & -          & -          & \checkmark          & -          & - \\
\texttt{COPELAND}   & -          & -          & -          & \checkmark          & \checkmark \\
\texttt{COPRAS}   & -          & -          & \checkmark          & \checkmark          & \checkmark \\
\texttt{CRADIS}   & -          & -          & -          & \checkmark          & \checkmark \\
\texttt{CRITIC}   & -          & -          & \checkmark          & \checkmark          & \checkmark \\
\texttt{DEMATEL}   & -          & -          & -          & \checkmark          & \checkmark \\
\texttt{EDAS}   & -          & -          & \checkmark          & \checkmark          & \checkmark \\
\texttt{ELECTRE}   & \checkmark          & -          & -          & \checkmark          & \checkmark \\
\texttt{ENTROPY}   & -          & -          & \checkmark          & \checkmark          & \checkmark \\
\texttt{GRA}   & -          & -          & -          & \checkmark          & \checkmark \\
\texttt{IDOCRIW}   & -          & -          & \checkmark          & \checkmark          & \checkmark \\
\texttt{MABAC}   & -          & -          & \checkmark          & \checkmark          & \checkmark \\
\texttt{MACBETH}   & -          & -          & -          & \checkmark          & \checkmark \\
\texttt{MAICRA}   & -          & -          & \checkmark          & \checkmark          & \checkmark \\
\texttt{MARCOS}   & -          & -          & \checkmark          & \checkmark          & \checkmark \\
\texttt{MARA}   & -          & -          & -          & \checkmark          & \checkmark \\
\texttt{MARE}   & \checkmark          & -          & -          & -          & - \\
\texttt{MAUT}   & -          & -          & -          & \checkmark          & \checkmark \\
\texttt{MEREC}   & -          & -          & \checkmark          & \checkmark          & \checkmark \\
\texttt{MOORA}   & -          & \checkmark          & \checkmark          & \checkmark          & \checkmark \\
\texttt{MOOSRA}   & -          & -          & -          & \checkmark          & \checkmark \\
\texttt{MRSORT}   & \checkmark          & -          & -          & -          & \checkmark \\
\texttt{MULTIMOORA}   & -          & \checkmark          & -          & \checkmark          & \checkmark \\
\texttt{OCRA}   & -          & -          & \checkmark          & \checkmark          & \checkmark \\
\texttt{ORESTE}   & -          & -          & -          & \checkmark          & \checkmark \\
\texttt{PIV}   & -          & -          & -          & \checkmark          & \checkmark \\
\texttt{PROMETHEE}   & \checkmark          & -          & \checkmark          & \checkmark          & \checkmark \\
\texttt{PSI}   & -          & -          & -          & \checkmark          & \checkmark \\
\texttt{REGIME}   & -          & -          & -          & \checkmark          & \checkmark \\
\texttt{RIM}   & -          & \checkmark          & -          & \checkmark          & \checkmark \\
\texttt{ROV}   & -          & -          & -          & \checkmark          & \checkmark \\
\texttt{SAW}   & -          & -          & -          & \checkmark          & \checkmark \\
\texttt{SBWM}   & -          & -          & -          & -          & \checkmark \\
\texttt{SECA}   & -          & -          & -          & \checkmark          & \checkmark \\
\texttt{SMART}   & -          & -          & -          & \checkmark          & \checkmark \\
\texttt{SMCDM}   & -          & -          & -          & -          & \checkmark \\
\texttt{SPOTIS}   & -          & -          & \checkmark          & \checkmark          & \checkmark \\
\texttt{SRMP}   & \checkmark          & -          & -          & -          & \checkmark \\
\texttt{SURE}   & \checkmark          & -          & -          & -          & - \\
\texttt{TODIM}   & -          & -          & -          & \checkmark          & \checkmark \\
\texttt{TOPSIS}   & \checkmark          & \checkmark          & \checkmark          & \checkmark          & \checkmark \\
\texttt{UTA}   & \checkmark          & -          & -          & -          & - \\
\texttt{VIKOR}   & \checkmark          & \checkmark          & \checkmark          & \checkmark          & \checkmark \\
\texttt{WASPAS}   & -          & \checkmark          & -          & \checkmark          & \checkmark \\
\texttt{WPM}   & -          & \checkmark          & -          & \checkmark          & \checkmark \\
\texttt{WSM}   & -          & \checkmark          & -          & \checkmark          & \checkmark \\
\texttt{WINGS}   & -          & -          & -          & \checkmark          & \checkmark \\
\texttt{WISP}   & -          & -          & -          & \checkmark          & \checkmark \\
\end{longtabu}

\section{Software description and architecture}
The \texttt{RMCDA} package can be divided into three main segments: (i) functions for reading input data from CSV or data frames, (ii) functions to apply a variety of MCDM methods (both classical and advanced), and (iii) visualization and reporting utilities. 

\subsection{Overview of key functions}
Table \ref{functionTab} lists and describes major functions of the RMCDA package.

\begin{table}[!h]
\centering
\caption{Key Functions in \texttt{RMCDA}}
\vspace{5mm}
\resizebox{\linewidth}{!}{%
\begin{tabular}{|c|c|c|}
\hline
\textbf{Function Name} & \textbf{Type} & \textbf{Description}  \\
\hline
\texttt{apply.AHP}         & MCDA method & Runs AHP on pairwise comparison data \\
\texttt{apply.ANP}         & MCDA method & Extends AHP to network structures \\
\texttt{apply.ARAS}         & MCDA method & Applies ARAS to data \\
\texttt{apply.BORDA}         & MCDA method & Applies BORDA to data \\
\texttt{apply.BWM}         & MCDA method & Applies Best-Worst Method \\
\texttt{apply.CILOS}         & Weighting & Finds criteria weights using CILOS\\
\texttt{apply.COCOSO}         & MCDA method & Applies COCOSO to data \\
\texttt{apply.CODAS}         & MCDA method & Applies CODAS to data \\
\texttt{apply.COPELAND}         & MCDA method & Applies COPELAND to data \\
\texttt{apply.COPRAS}         & MCDA method & Applies COPRAS to data \\
\texttt{apply.CRADIS}         & MCDA method & Applies CRADIS to data \\
\texttt{apply.CRITIC}      & Weighting    & Finds criteria weights using CRITIC \\
\texttt{apply.DEMATEL}         & Weighting & Finds criteria weights using DEMATEL \\
\texttt{apply.EDAS}         & MCDA method & Applies EDAS to data \\
\texttt{apply.ELECTRE1}     & MCDA method    & Applies ELECTRE I method to data \\
\texttt{apply.Entropy}     & Weighting    & Finds criteria weights using Entropy \\
\texttt{apply.GRA}      & MCDA method & Applies GRA to data \\
\texttt{apply.IDOCRIW}      & Weighting & Finds criteria weights using IDOCRIW \\
\texttt{apply.MABAC}      & MCDA method & Applies MABAC to data \\
\texttt{apply.MACBETH}      & MCDA method & Applies MACBETH to data \\
\texttt{apply.MAIRCA}      & MCDA method & Applies MAIRCA to data \\
\texttt{apply.MARA}      & MCDA method & Applies MARA to data \\
\texttt{apply.MAUT}      & MCDA method & Applies MAUT to data \\
\texttt{apply.MARCOS}      & MCDA method & Applies MARCOS to data \\
\texttt{apply.MOORA}      & MCDA method & Applies MOORA to data \\
\texttt{apply.MOOSRA}      & MCDA method & Applies MOOSRA to data \\
\texttt{apply.MULTIMOORA}      & MCDA method & Applies Multi-MOORA to data \\
\texttt{apply.OCRA}      & MCDA method & Applies OCRA to data \\
\texttt{apply.OPA}      & MCDA method & Applies OPA to data \\
\texttt{apply.ORESTE}      & MCDA method & Applies ORESTE to data \\
\texttt{apply.PIV}      & MCDA method & Applies PIV to data \\
\texttt{apply.PROMETHEE}      & MCDA method & Applies PROMETHEE to data \\
\texttt{apply.PSI}      & MCDA method & Applies PSI to data \\
\texttt{apply.RAFSI}      & MCDA method & Applies RAFSI to data \\
\texttt{apply.REGIME}      & MCDA method & Applies REGIME to data \\
\texttt{apply.RIM}      & MCDA method & Applies RIM to data \\
\texttt{apply.ROV}      & MCDA method & Applies ROV to data \\
\texttt{apply.SAW}      & MCDA method & Applies SAW to data \\
\texttt{apply.SBWM}        & MCDA method & Applies Stratified Best-Worst Method to data \\
\texttt{apply.SECA}      & Weighting & Finds criteria weights using SECA \\
\texttt{apply.SMART}      & MCDA method& Applies SMART to data \\
\texttt{apply.SMCDM}       & MCDA method & Stratified MCDM solution \\
\texttt{apply.SPOTIS}      & MCDA method& Applies SPOTIS to data \\
\texttt{apply.SRMP}       & MCDA method & Applies SRMP to data  \\
\texttt{apply.TODIM}       & MCDA method & Applies TODIM to data  \\
\texttt{apply.TOPSIS}       & MCDA method & Applies TOPSIS to data  \\
\texttt{apply.VIKOR}       & MCDA method & Applies VIKOR to data\\
\texttt{apply.WASPAS}       & MCDA method & Applies WASPAS to data \\
\texttt{apply.WINGS}       & Weighting & Finds criteria weights using WINGS \\
\texttt{apply.WISP}       & MCDA method & Applies WISP to data \\
\texttt{apply.WSM}       & MCDA method & Applies WSM to data \\
\texttt{apply.WPM}       & MCDA method & Applies WPM to data \\

\hline
\texttt{read.csv.AHP.matrices}, \texttt{read.csv.SBWM.matrices}, \ldots & I/O & Import specialized CSV formats \\
\hline
\texttt{plot.AHP.decision.tree}, \texttt{plot.spider}, etc. & Visualization & Decision tree and radar/spider charts \\
\hline \label{functionTab}
\end{tabular}
}
\end{table}

\subsection{Software functionalities}
RMCDA offers a range of MCDM methods, each implemented with user-friendly functionality and optional visualizations to enhance interpretability of data. Below we provide an overview of three key methods that were not previously included in any comprehensive MCDA packages:

\subsection{Analytical Network Process (ANP)}
The Analytic Network Process (ANP) extends the Analytic Hierarchy Process (AHP) introduced by Saaty in 1980 by incorporating network structures rather than simple hierarchies \cite{saaty1980analytic}. ANP evaluates dependencies and feedback loops between criteria and alternatives. The methodology involves performing pairwise comparisons and computing priority weights using eigenvector analysis \cite{saaty2006decision}. ANP first extracts the criteria weight vector and the weighted matrix for alternatives. Then, it constructs a supermatrix that integrates these values, organizing the structure into a larger matrix where decision elements interact. This includes placing the criteria weights, the alternatives' weighted values, and an identity matrix to account for dependencies between alternatives. Finally, the function raises the supermatrix to a specified power, amplifying the influence of network relationships, and returns the processed matrix. This refined structure captures both direct and indirect dependencies in decision-making, making ANP a more robust method for complex evaluations compared to AHP alone.

\subsection{Stratified MCDM (SMCDM)}
\begin{figure}[ht]
    \centering
    \includegraphics[width=1\linewidth]{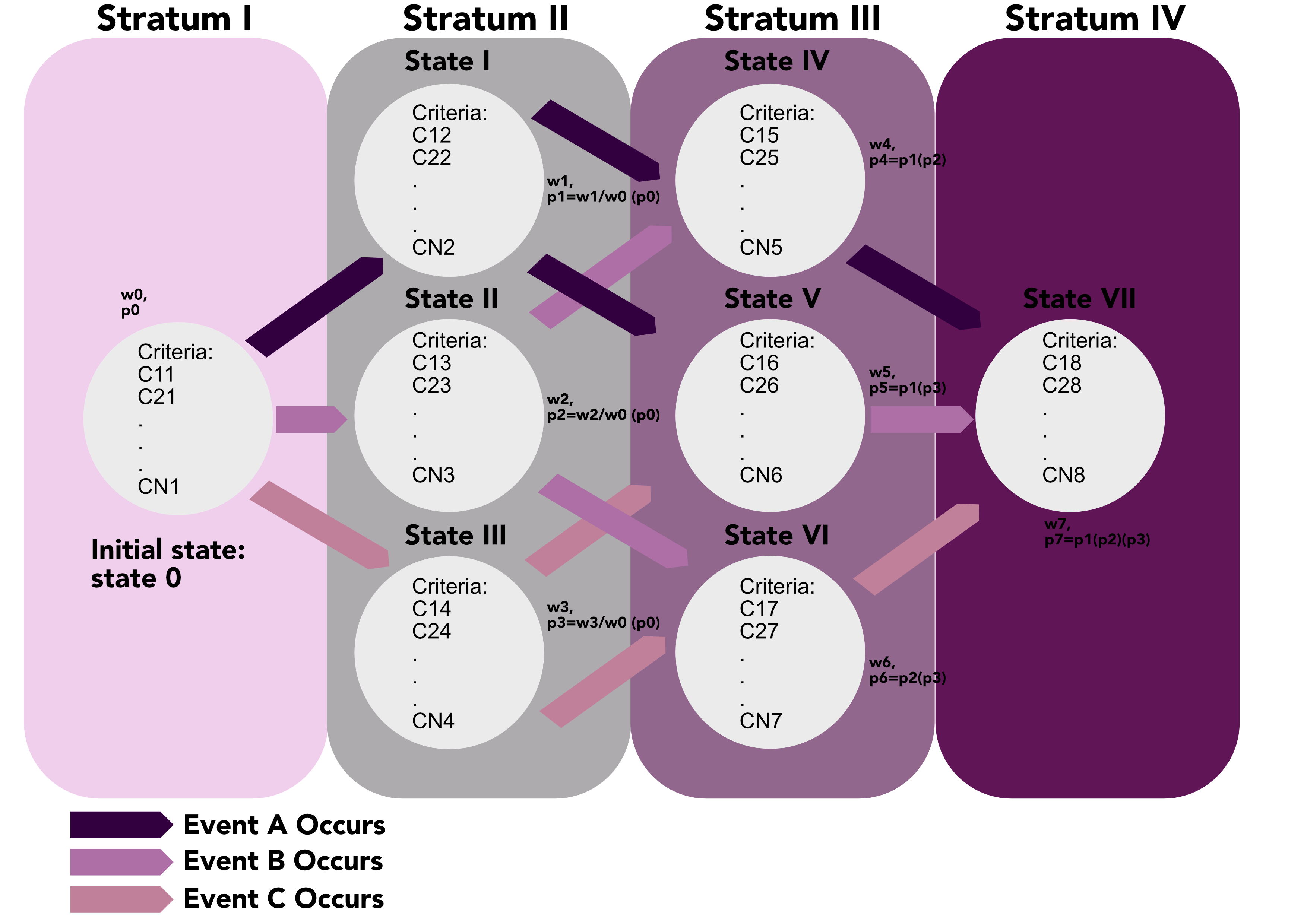}
    \caption{\textbf{Flowchart demonstrating how the strata and states are structured in the SMCDM method with three events which occur independent of each other.}}
    \label{SMCDM-stratums}
\end{figure}

The Stratified MCDM (SMCDM) method was developed by Asadabadi in 2018 \cite{asadabadi2018stratified} and is useful for situations involving uncertainty such as a buying a house where the importance of criteria (price, number of rooms, or proximity to schools) may change depending on the customer's situation. For example, the customer may earn a promotion with some probability which affects the importance of the price of the house. We consider the initial situation of the customer as state 0 located in stratum I and create a second stratum (stratum II) which contains states with the occurrence of only one possible event. In Figure \ref{SMCDM-stratums} we show three possible events that may happen; event $A$, event $B$, and event $C$. We create a third stratum (stratum III) which contains states with the concurrent occurrence of two states together and lastly we create a fourth stratum with one state in which all three states happen simultaneously. If the probabilities of occurrence of each state across strati are given then we calculate the weights of each criterion then the scores of alternatives. Otherwise, if then events happen independently of each other and the probability of occurrence of states in strati I and II are given, then we calculate the probability of occurrence of each state with respect to the first probability of staying in the baseline state (state 0). We know that the sum of the probabilities of the system being in either of the states should equate 1. Therefore:

\begin{equation}
    \sum_{i =0}^{8}p_i = p_0 + (\frac{w_1}{w_0}+\frac{w_2}{w_0}+\frac{w_3}{w_0})p_0 + (\frac{w_1}{w_0}\frac{w_2}{w_0} + \frac{w_1}{w_0}\frac{w_3}{w_0} + \frac{w_2}{w_0}\frac{w_3}{w_0})p_0^2 + (\frac{w_1}{w_0}\frac{w_2}{w_0}\frac{w_3}{w_0})p_0^3 = 1
\end{equation}
$p_i$ and $w_j$ in the above equation represent the probability of occurrence of state $i$ and the associated weight of the event $j$ respectively.
Our R package provides an \texttt{apply.SMCDM} function which automatically sets up this problem, solves this equation and finds the optimal option based on the non-imaginary root of this polynomial equation in the case where we only know the probability of occurrence of single independent events. We provide detailed instructions regarding the use of this function in the Example II section. 
\subsection{Stratified Best Worst Method (SBWM)}
The SBWM method calculates optimal decision values for various alternatives across different states based on specified criteria. The method requires several inputs: a comparison matrix of alternatives and criteria, matrices for comparing criteria to the `worst' and `best' criteria in each state, lists specifying the worst and best criteria per state, and a vector of likelihoods indicating the probability of each state. The methodology involves iterating over each state and calculating the weight vector for each state's criteria. The resulting state-specific weights are compiled into a matrix and processed along with the likelihood vector following the same steps as the SMCDM method. The output is a numerical result representing the optimal decision values for the alternatives based on the weighted criteria across states \cite{torkayesh2021sustainable}.
\subsection{ShinyRMCDA}
\begin{figure}
    \centering
    \includegraphics[width=1\linewidth]{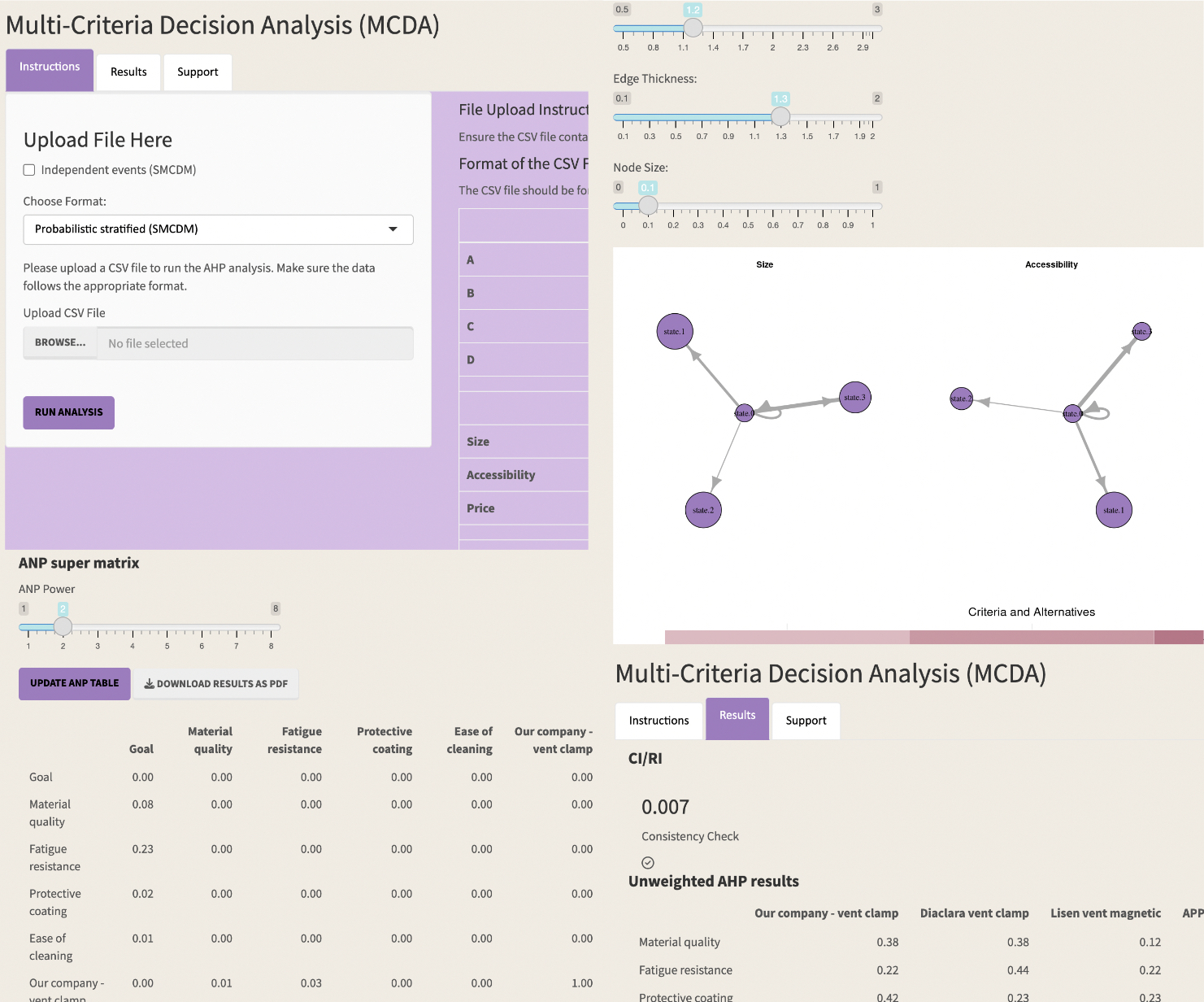}
    \caption{\textbf{Overview of the ShinyRMCDA application.}}
    \label{shinyRMCDA-overview}
\end{figure}

To provide easier accessibility to these methodologies to users with no programming experience, we have developed a web-based application which receives a CSV file as input and provides tools for visualization and data analysis. To use the application please use the following link: \href{https://najafiannice.shinyapps.io/AHP_app/}{shinyRMCDA}. Upon visiting the web-based application, you will be guided through an instructions page where you can select a methodology from the drop-down menu. The instructions page would get automatically updated upon selection of a methodology. After running the analysis, the application outputs a set of plots and tables which can be easily downloaded from the web page (Figure \ref{shinyRMCDA-overview} provides an overview of the ShinyRMCDA web-based application). 
\section{Illustrative examples}
\subsection{Installation}
\begin{enumerate}
\item Ensure you have \texttt{R} version $>$= 4.0 installed.
\item Install the required dependencies, e.g.:
\begin{lstlisting}
install.packages(c("devtools", "igraph", "lpSolve", "fmsb"))
\end{lstlisting}
\item Install \texttt{RMCDA} from GitHub:
\begin{lstlisting}
devtools::install_github("AnniceNajafi/RMCDA")
\end{lstlisting}
\end{enumerate}
\textbf{Note:} If you encounter an error that \texttt{devtools} or \texttt{igraph} is not available, please install them manually via \texttt{install.packages("devtools")} or \texttt{install.packages("igraph")}.
\subsection{Example I.} 
\begin{figure}
    \centering
    \includegraphics[width=0.9\linewidth]{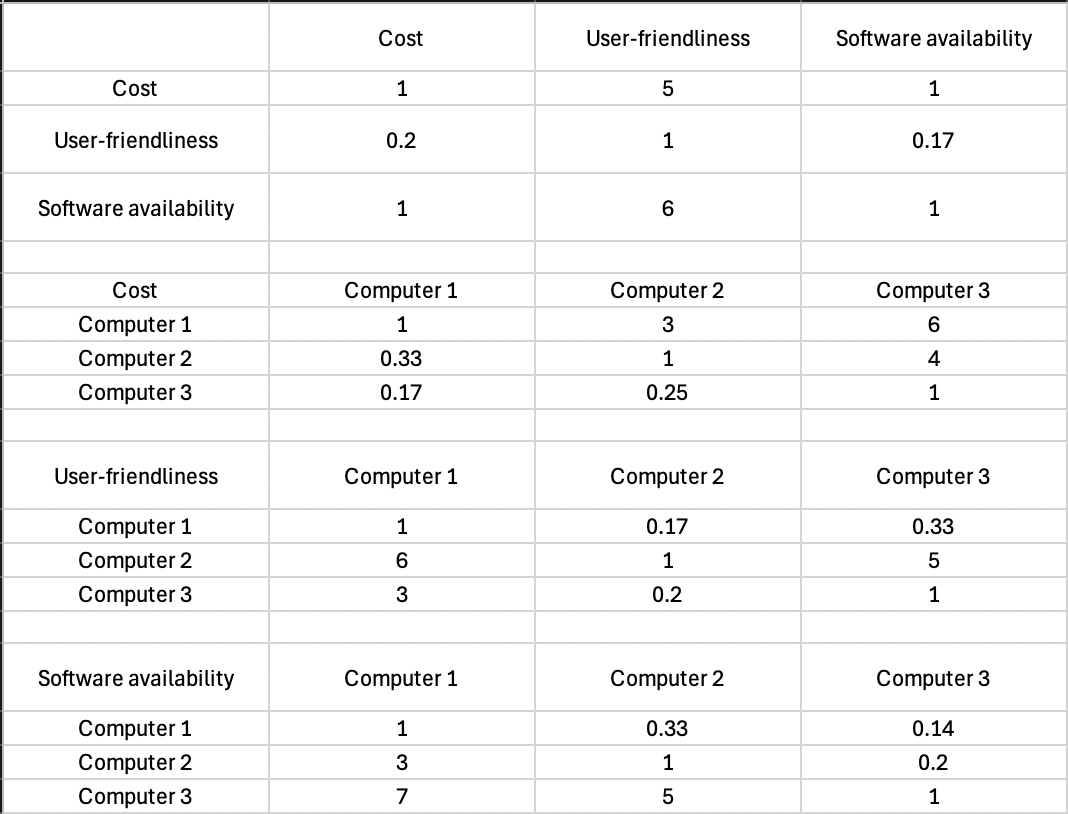}
    \caption{\textbf{Format of the input CSV file for AHP.}}
    \label{CSV-input-AHP}
\end{figure}

\begin{figure}
    \centering
    \includegraphics[width=1.1\linewidth]{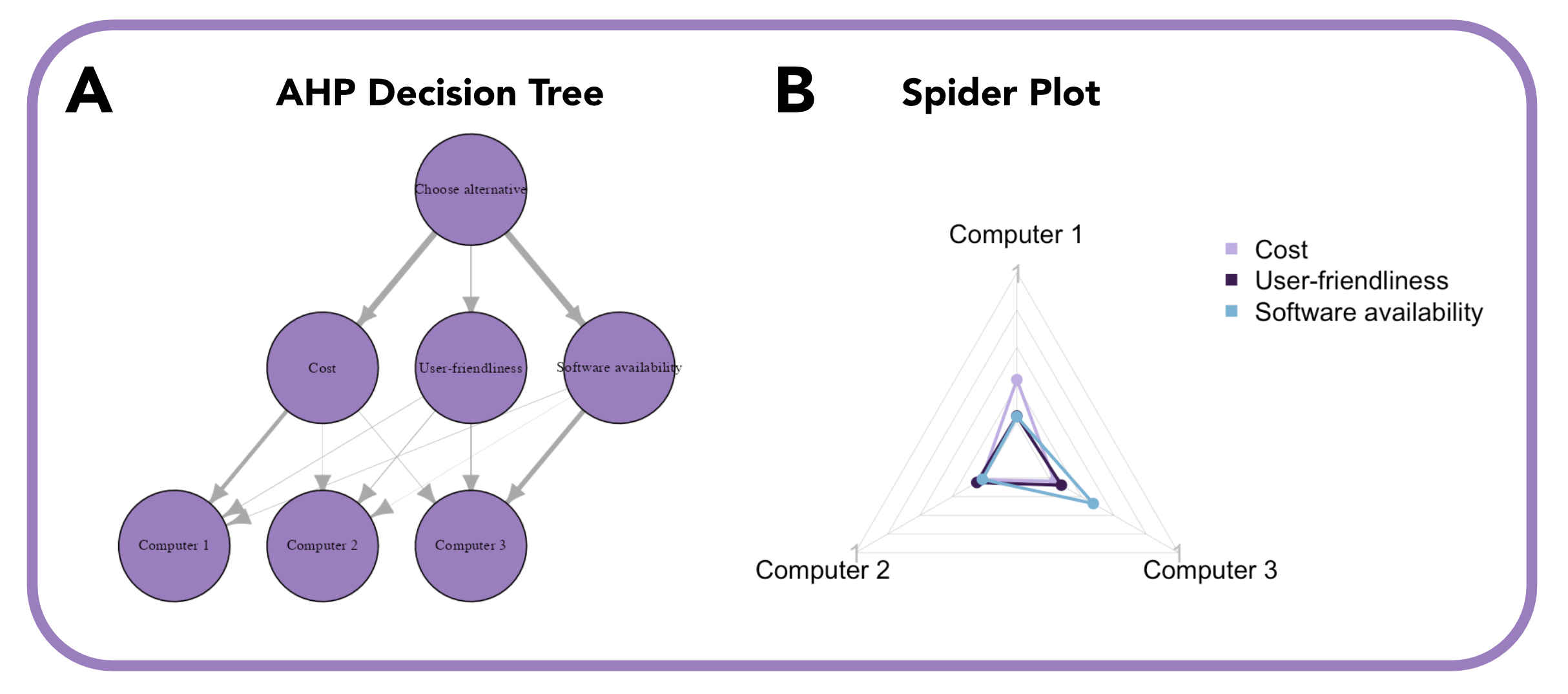}
    \caption{\textbf{Visualization outputs from the RMCDA package for AHP.}}
    \label{AHPDecisionTree}
\end{figure}

We are deciding to purchase a computer and we have compared the computers in a pairwise fashion based on three criteria; cost, user friendliness, and software availability. In addition, we have also compared the importance of the criteria to each other in a pairwise fashion. We have stored the results in a CSV file as shown in Figure \ref{CSV-input-AHP}. The CSV file should contain the inputs for the AHP method in a sequential order starting with the pairwise comparison of criteria followed by matrices related to pairwise comparison of alternatives for each criterion. The number of inputs in the CSV file would always be $n+1$ with $n$ being the number of criteria.
Our goal is to find out which computer we should buy. Below we demonstrate how we can utilize the RMCDA package to load the parameters stored in a CSV file and apply AHP and ANP to the data. 
\begin{lstlisting}
>library(RMCDA)
>data <- read.csv("AHP_example_computers.csv", header=FALSE)
>data.lst <- read.csv.AHP.matrices(data)
>A <- data.lst[[1]]
>comparing.competitors <- data.lst[[2]]
>AHP.result <- apply.AHP(A, comparing.competitors)
\end{lstlisting}
The function returns four outputs. The first output is the ratio of the consistency index to the random index where a value below $0.1$ indicates no significant inconsistencies found in data. The second and third outputs return the unweighted and weighted scores matrix respectively. Lastly, the fourth output returns the final AHP scores for the three alternatives. To visualize the decision tree corresponding to the AHP process, we can use the following code:
\begin{lstlisting}
>plot.AHP.decision.tree(A, comparing.competitors)
\end{lstlisting}
We can also use the \texttt{plot.spider} function to show the scores or weights of criteria as a radar plot. Figure \ref{AHPDecisionTree} A and B depict the AHP decision tree and the radar plot corresponding to the weighted matrix for criteria related to this example respectively (The same figure can be generated if the input CSV file is uploaded in ShinyRMCDA). 

To apply ANP on the data, we can utilize the \texttt{apply.ANP} function which will output the supermatrix of the ANP method. 

\begin{lstlisting}
>power = 3 #exponential for the super matrix. 
>apply.ANP(A, comparing.competitors, power)
\end{lstlisting}

\subsection{Example II.}
\begin{figure}
    \centering
    \includegraphics[width=0.5\linewidth]{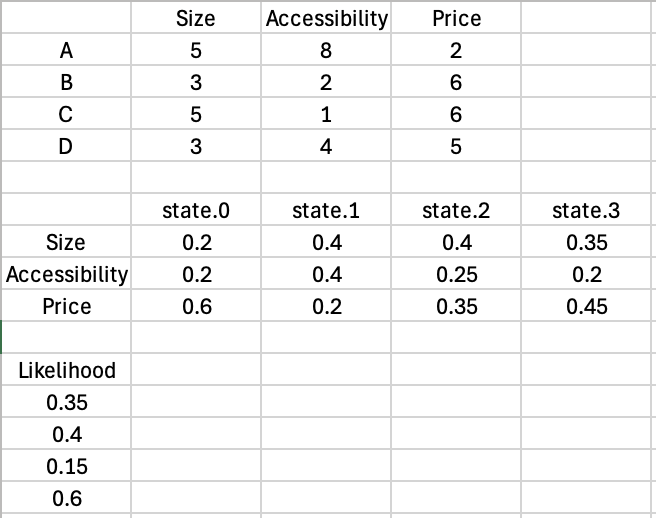}
    \caption{\textbf{The input CSV file for the SMCDM methodology.}}
    \label{SMCDM-CSV}
\end{figure}

\begin{figure}
    \centering
    \includegraphics[width=1\linewidth]{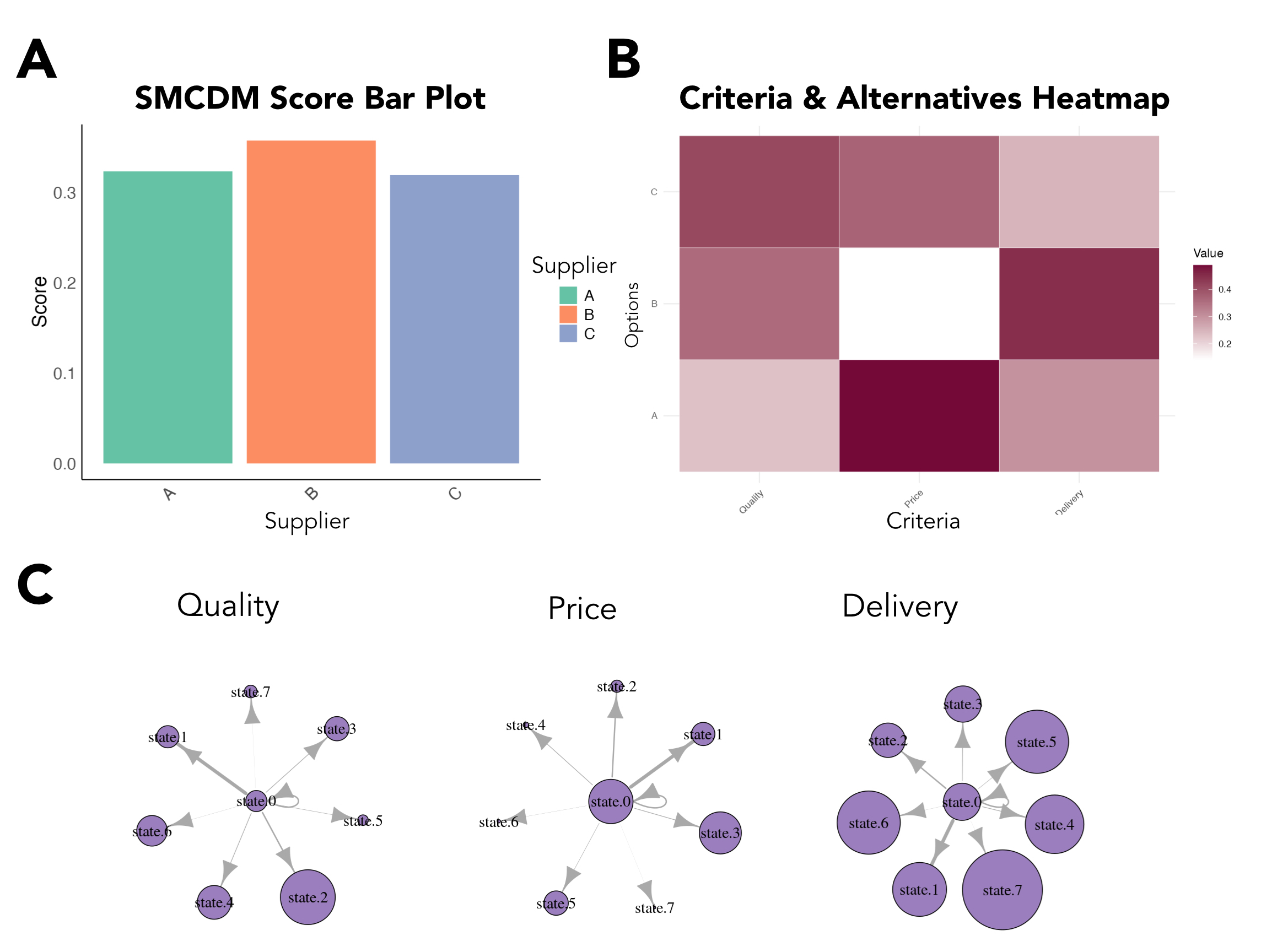}
    \caption{\textbf{Visualization outputs from the RMCDA package for SMCDM.}}
    \label{fig:SMCDM-shiny-output}
\end{figure}

Suppose we are buying a house and we have specific criteria based on which we weigh our options. At our current situation we have ranked our criteria in Table \ref{AlternativesSMCDM} below.
\begin{table}[h!]
\centering
\caption{\textbf{Scores of alternatives based on criteria.} }
\vspace{5mm}
\begin{tabular}{|c|c|c|c|}
\hline
\textbf{Alternatives} & \textbf{Quality} & \textbf{Price} & \textbf{Delivery} \\
\hline
A & 0.23 & 0.49 & 0.3 \\
\hline
B & 0.36 & 0.14 & 0.45 \\
\hline
C & 0.41 & 0.37 & 0.25 \\
\hline
\end{tabular}
\label{AlternativesSMCDM}
\end{table}
With probability $0.17$ we will stay in our current state and with probabilities $0.42, 0.17, 0.08$ events $A, B$, and $C$ may happen and we would switch from our current state (state 0) to one of seven states with states 1 to 3 representing the occurrence of only one event, states 4 to 6 representing the concurrent occurrence of two events (happening with probabilities $0.08, 0.05, 0.02$ respectively and state 7 representing the occurrence of all events together. We are aware that the importance of criteria in states are as shown in Table \ref{TableIII}. The goal is to apply SMCDM to our data and score each alternative based in this particular example involving uncertainty. 
\begin{table}[h!]
\centering
\caption{\textbf{Scores of alternatives based on criteria}}
\vspace{5mm}
\begin{tabular}{|c|c|c|c|c|c|c|c|c|}
\hline
\textbf{Criteria} & \textbf{State 0} & \textbf{State I} & \textbf{State II} & \textbf{State III} & \textbf{State IV} & \textbf{State V} & \textbf{State VI} & \textbf{State VII}\\
\hline
Quality & 0.2 & 0.21 & 0.52 & 0.23 & 0.32 & 0.1 & 0.29 & 0.12\\
\hline
Price & 0.4 & 0.21 & 0.11 & 0.38 & 0.05 & 0.22 & 0.03 & 0.02\\
\hline
Delivery & 0.4 & 0.58 & 0.37 & 0.39 & 0.63 & 0.68 & 0.68 & 0.86\\
\hline
\end{tabular}
\label{TableIII}
\end{table}
We can either define the variables manually or store them in a CSV file as shown in Figure \ref{SMCDM-CSV} and read and process them using the following code:
\begin{lstlisting}
>data <- read.csv("SMCDM_input.csv", header=FALSE)
>data.lst <- read.csv.SMCDM.matrices(data)
>comparison.mat <- data.lst[[1]] 
>state.criteria.probs<- data.lst[[2]] 
>likelihood.vector <- data.lst[[3]]
>apply.SMCDM(comparison.mat, state.criteria.probs, likelihood.vector, independent = FALSE)
\end{lstlisting}
The CSV file should contain the comparison matrix for the alternatives based on different criteria followed by the importance of different criteria in each state and the likelihood of staying in the baseline state and occurrence of each state. 
In Figure \ref{fig:SMCDM-shiny-output} we show the output of the shinyRMCDA app for the same CSV file. Figure \ref{fig:SMCDM-shiny-output} A  shows the barplot corresponding to the scores of options based found through SMCDM, and B illustrates the heatmap for the importance of each criterion related to each alternative. C shows the decision tree with edge widths corresponding to the likelihood of transitioning to each state and the node size demonstrating the importance of the criteria in each state.
\subsection{Example III.}

\begin{figure}
    \centering
    \includegraphics[width=0.9\linewidth]{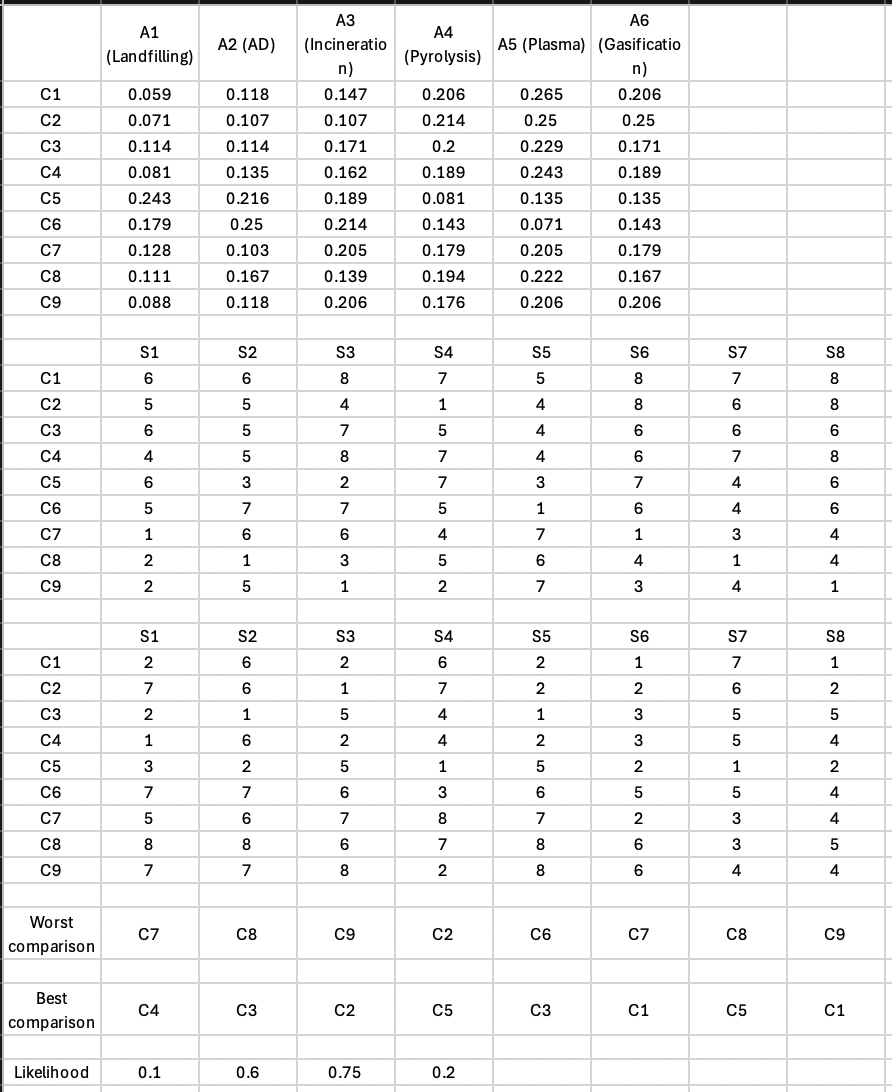}
    \caption{\textbf{Format of the CSV file containing data for the application of SBWM.}}
    \label{Torkayesh-csv}
\end{figure}

For this example we use the first case study data from Torkayesh et al \cite{torkayesh2021sustainable}. We have stored the inputs of the case study in a CSV file in the format shown in Figure \ref{Torkayesh-csv}. The first input is a table containing the importance of each criteria corresponding to each alternative. The table is then followed by two tables containing information related to the rankings of the worst criteria to others and the best criteria to others for every state respectively. We have then included two vectors indicating the worst and best criteria in each state respectively. Lastly, we have stored the likelihood of the occurrence of each event in the CSV file.

To apply SBWM we utilize the following code:
\begin{lstlisting}
>df.path <- read.csv(input$file$datapath, header=FALSE)
>df.lst <- read.csv.SBWM.matrices(df.path)
>comparison.mat <- df.lst[[1]]
>others.to.worst <-  df.lst[[2]]
>others.to.best <- df.lst[[3]]
>state.worst.lst <- df.lst[[4]]
>state.best.lst <- df.lst[[5]]
>likelihood.vector <- df.lst[[6]]
>SBWM.results <- SBWM(comparison.mat, others.to.worst, others.to.best, >state.worst.lst, state.best.lst, likelihood.vector)[,1]
\end{lstlisting}

\section{Algorithm Run Time}
\begin{figure}
    \centering
    \includegraphics[width=0.9\linewidth]{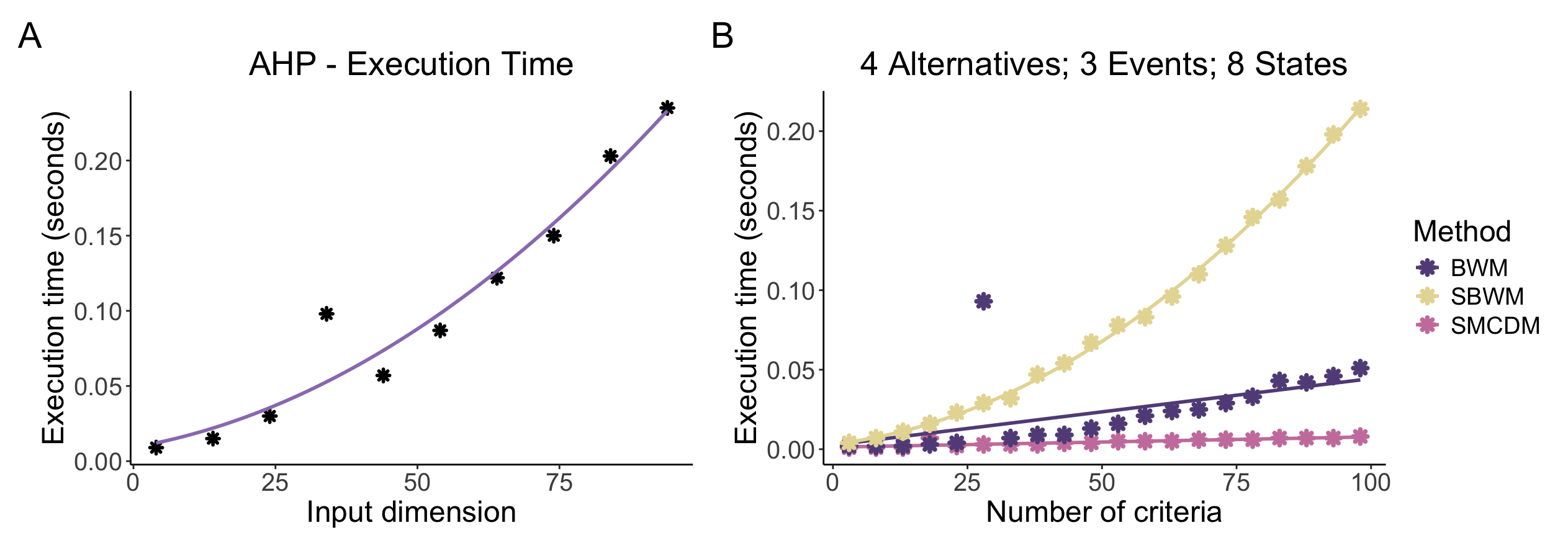}
    \caption{\textbf{Time complexity of the BWM, SBWM, and SMCDM methods compared to the AHP method.}}
    \label{figAlgorithmRunTimes}
\end{figure}

We generated random inputs for the BWM, SMCDM, and SBWM functions, each time increasing the number of criteria and recorded the time it takes to run the functions. We then fitted polynomial equations to the curves which demonstrated that the time complexity of the BWM and SMCDM are both linear ($O(n)$) while the time complexity of the SBWM method is quadratic ($O(n^2)$) as shown in Figure \ref{figAlgorithmRunTimes} B. For the purpose of comparison, we have plotted the time complexity of running the AHP algorithm on data and have shown the results in Figure \ref{figAlgorithmRunTimes} A demonstrating that the time complexity of the SBWM algorithm with 3 independent events is similar to the AHP method.
\section{Impact}

Despite the increasing interest in Multi-Criteria Decision Analysis (MCDA), R packages offer only a limited number of MCDA methods, and their usage in R has been declining. We established RMCDA to fulfill the need for MCDA methods inside the R ecosystem by providing comprehensive coverage of popular MCDA methodologies. RMCDA integrates a wide array of classical, probabilistic, and stratified approaches into a single package, thereby exceeding the functionalities of current R packages and reviving R's status as a platform for MCDA applications. Additionally, by using the user-friendly online application ShinyRMCDA, we extend the accessibility of these methodologies for academics and practitioners with minimal programming expertise, hence promoting wider acceptance and implementation in decision-making scenarios. This program strengthens computational decision analysis and fits with the growing multidisciplinary dependence on MCDA approaches across many fields, including healthcare, finance, engineering, and environmental planning.
\section{Conclusions}
Despite the growing demand for MCDA methodologies, current R packages offer limited implementations, resulting in a decrease in their utilization. On the other hand, Python has experienced an increase in MCDA-related libraries and their use, suggesting a demand for a strong alternative in R. Here, we introduced a comprehensive R package, RMCDA that offers a user-friendly MCDA framework. RMCDA encompasses a wide range of traditional, probabilistic, and stratified decision-making techniques, while also improving accessibility via its ShinyRMCDA web application, which is accessible to users with limited programming skills.

Although RMCDA offers comprehensive capabilities, some limitations remain. At present, it excludes fuzzy MCDA methods, which are essential for decision-making in conditions of uncertainty or imprecision. Future versions of RMCDA will integrate fuzzy logic techniques. Moreover, computational efficiency will be enhanced for large-scale applications, especially concerning high-dimensional datasets. Despite these constraints, RMCDA serves as decision-support tool for researchers and practitioners in various sectors, including healthcare, finance, engineering, and environmental planning who utilize the R programming language.


\printbibliography

@article{mardani2015multiple,
  title={Multiple criteria decision-making techniques and their applications--a review of the literature from 2000 to 2014},
  author={Mardani, Abbas and Jusoh, Ahmad and Nor, Khalil and Khalifah, Zainab and Zakwan, Norhayati and Valipour, Alireza},
  journal={Economic research-Ekonomska istra{\v{z}}ivanja},
  volume={28},
  number={1},
  pages={516--571},
  year={2015},
  publisher={Sveu{\v{c}}ili{\v{s}}te Jurja Dobrile u Puli, Odjel za ekonomiju i turizam'Dr. Mijo~…}
}

@book{saaty2006decision,
  title={Decision making with the analytic network process},
  author={Saaty, Thomas L and Vargas, Luis G and others},
  volume={282},
  year={2006},
  publisher={Springer}
}

@article{rezaei2015best,
  title={Best-worst multi-criteria decision-making method},
  author={Rezaei, Jafar},
  journal={Omega},
  volume={53},
  pages={49--57},
  year={2015},
  publisher={Elsevier}
}

@article{asadabadi2018stratified,
  title={The stratified multi-criteria decision-making method},
  author={Asadabadi, Mehdi Rajabi},
  journal={Knowledge-Based Systems},
  volume={162},
  pages={115--123},
  year={2018},
  publisher={Elsevier}
}

@article{zionts1979mcdm,
  title={MCDM—If not a roman numeral, then what?},
  author={Zionts, Stanley},
  journal={Interfaces},
  volume={9},
  number={4},
  pages={94--101},
  year={1979},
  publisher={INFORMS}
}

@Manual{AHPtools,
    title = {AHPtools: Consistency in the Analytic Hierarchy Process},
    author = {Amarnath Bose},
    year = {2024},
    note = {R package version 0.3.0},
    url = {https://CRAN.R-project.org/package=AHPtools},
}

@article{saaty1980analytic,
  title={The analytic hierarchy process (AHP)},
  author={Saaty, Thomas L},
  journal={The Journal of the Operational Research Society},
  volume={41},
  number={11},
  pages={1073--1076},
  year={1980}
}

@Manual{lpSolve,
    title = {lpSolve: Interface to 'Lp_solve' v. 5.5 to Solve Linear/Integer Programs},
    author = {Michel Berkelaar},
    year = {2024},
    note = {R package version 5.6.21},
    url = {https://CRAN.R-project.org/package=lpSolve},
}

@article{torkayesh2021sustainable,
  title={Sustainable waste disposal technology selection: The stratified best-worst multi-criteria decision-making method},
  author={Torkayesh, Ali Ebadi and Malmir, Behnam and Asadabadi, Mehdi Rajabi},
  journal={Waste Management},
  volume={122},
  pages={100--112},
  year={2021},
  publisher={Elsevier}
}

@article{bigaret2017supporting,
  title={Supporting the multi-criteria decision aiding process: R and the MCDA package},
  author={Bigaret, S{\'e}bastien and Hodgett, Richard E and Meyer, Patrick and Mironova, Tatiana and Olteanu, Alexandru-Liviu},
  journal={EURO journal on decision processes},
  volume={5},
  number={1-4},
  pages={169--194},
  year={2017},
  publisher={Elsevier}
}

@article{diakoulaki1995determining,
  title={Determining objective weights in multiple criteria problems: The critic method},
  author={Diakoulaki, Danae and Mavrotas, George and Papayannakis, Lefteris},
  journal={Computers \& Operations Research},
  volume={22},
  number={7},
  pages={763--770},
  year={1995},
  publisher={Elsevier}
}

@book{tzeng2011multiple,
  title={Multiple attribute decision making: methods and applications},
  author={Tzeng, Gwo-Hshiung and Huang, Jih-Jeng},
  year={2011},
  publisher={CRC press}
}

@article{rocha2024review,
  title={Review and Bibliographic Analysis of Metaheuristic Methods in Multicriteria Decision-Making: A 45-Year Perspective Across International, Latin American, and Colombian Contexts},
  author={Rocha, Christian Manuel Moreno and Benitez, Andres Santacruz and Buelvas, Daina Arenas},
  journal={Journal of Applied Mathematics},
  volume={2024},
  number={1},
  pages={5577682},
  year={2024},
  publisher={Wiley Online Library}
}

@article{brans2016promethee,
  title={PROMETHEE methods},
  author={Brans, Jean-Pierre and De Smet, Yves},
  journal={Multiple criteria decision analysis: state of the art surveys},
  pages={187--219},
  year={2016},
  publisher={Springer}
}

@article{srivastava2024multi,
  title={Multi-criteria decision making (MCDM) in diverse domains of education: a comprehensive bibliometric analysis for research directions},
  author={Srivastava, Shilpa and Tripathi, Aprna and Arora, Nidhi},
  journal={International Journal of System Assurance Engineering and Management},
  pages={1--18},
  year={2024},
  publisher={Springer}
}

@Manual{topsisR,
    title = {topsis: TOPSIS method for multiple-criteria decision making (MCDM)},
    author = {Mahmoud Mosalman Yazdi},
    year = {2013},
    note = {R package version 1.0},
    url = {https://CRAN.R-project.org/package=topsis}
}

@Manual{ahpsurveyR,
    title = {ahpsurvey: Analytic Hierarchy Process for Survey Data},
    author = {Frankie Cho},
    year = {2019},
    note = {R package version 0.4.1},
    url = {https://CRAN.R-project.org/package=ahpsurvey},
}

@Manual{MCDAR,
    title = {MCDA: Support for the Multicriteria Decision Aiding Process},
    author = {Patrick Meyer and Sébastien Bigaret and Richard Hodgett and Alexandru-Liviu Olteanu and David Palma and Aritad Alan Choicharoon},
    year = {2023},
    note = {R package version 0.1.0},
    url = {https://CRAN.R-project.org/package=MCDA},
}

@article{ishizaka2018visual,
  title={Visual management of performance with PROMETHEE productivity analysis},
  author={Ishizaka, Alessio and Resce, Giuliano and Mareschal, Bertrand},
  journal={Soft Computing},
  volume={22},
  pages={7325--7338},
  year={2018},
  publisher={Springer}
}

@article{greco2019methodological,
  title={On the methodological framework of composite indices: A review of the issues of weighting, aggregation, and robustness},
  author={Greco, Salvatore and Ishizaka, Alessio and Tasiou, Menelaos and Torrisi, Gianpiero},
  journal={Social indicators research},
  volume={141},
  pages={61--94},
  year={2019},
  publisher={Springer}
}

@article{pereira2024enhancing,
  title={Enhancing decision analysis with a large language model: pydecision a comprehensive library of MCDA methods in python},
  author={Pereira, Valdecy and Basilio, Marcio Pereira and Santos, Carlos Henrique Tarjano SantosCarlos Henrique Tarjano},
  journal={arXiv preprint arXiv:2404.06370},
  year={2024}
}

@article{kizielewicz2023pymcdm,
  title={pymcdm—The universal library for solving multi-criteria decision-making problems},
  author={Kizielewicz, Bart{\l}omiej and Shekhovtsov, Andrii and Sa{\l}abun, Wojciech},
  journal={SoftwareX},
  volume={22},
  pages={101368},
  year={2023},
  publisher={Elsevier}
}

@article{wkatrobski2022pyrepo,
  title={pyrepo-mcda—Reference objects based MCDA software package},
  author={W{\k{a}}tr{\'o}bski, Jaros{\l}aw and B{\k{a}}czkiewicz, Aleksandra and Sa{\l}abun, Wojciech},
  journal={SoftwareX},
  volume={19},
  pages={101107},
  year={2022},
  publisher={Elsevier}
}

@article{satman2021jmcdm,
  title={JMcDM: A Julia package for multiple-criteria decision-making tools},
  author={Satman, Mehmet Hakan and Y{\i}ld{\i}r{\i}m, Bahad{\i}r Fatih and Kuruca, Ersagun},
  journal={The Journal of Open Source Software},
  volume={6},
  number={65},
  pages={1--6},
  year={2021}
}

@article{meyer2012diviz,
  title={Diviz: A software for modeling, processing and sharing algorithmic workflows in MCDA},
  author={Meyer, Patrick and Bigaret, S{\'e}bastien},
  journal={Intelligent decision technologies},
  volume={6},
  number={4},
  pages={283--296},
  year={2012},
  publisher={IOS Press}
}

@Manual{pkgsearch,
    title = {pkgsearch: Search and Query CRAN R Packages},
    author = {Gábor Csárdi and Maëlle Salmon},
    year = {2025},
    note = {R package version 3.1.4},
    url = {https://CRAN.R-project.org/package=pkgsearch},
}

@article{levwilk,
  title={Wilk, J. pypinfo: View PyPI download statistics with ease (2018)},
  author={Lev, O and Dufresne, J and Kasim, R and Skinn, B},
  journal={URL https://github. com/ofek/pypinfo}
}

@Manual{cranlogs,
    title = {cranlogs: Download Logs from the 'RStudio' 'CRAN' Mirror},
    author = {Gábor Csárdi},
    year = {2019},
    note = {R package version 2.1.1},
    url = {https://CRAN.R-project.org/package=cranlogs},
}

@article{saaty2004decision,
  title={Decision making—the analytic hierarchy and network processes (AHP/ANP)},
  author={Saaty, Thomas L},
  journal={Journal of systems science and systems engineering},
  volume={13},
  pages={1--35},
  year={2004},
  publisher={Springer}
}

@incollection{pehlivan2017comparison,
  title={Comparison of methods in FAHP with application in supplier selection},
  author={Pehlivan, Nimet Yapici and Paksoy, Turan and {\c{C}}alik, Ahmet},
  booktitle={Fuzzy analytic hierarchy process},
  pages={45--76},
  year={2017},
  publisher={Chapman and Hall/CRC}
}

@article{zavadskas2010new,
  title={A new additive ratio assessment (ARAS) method in multicriteria decision-making},
  author={Zavadskas, Edmundas Kazimieras and Turskis, Zenonas},
  journal={Technological and economic development of economy},
  volume={16},
  number={2},
  pages={159--172},
  year={2010},
  publisher={Taylor \& Francis}
}

@article{borda1781m,
  title={M'emoire sur les' elections au scrutin},
  author={Borda, JC de},
  journal={Histoire de l'Acad'emie Royale des Sciences},
  year={1781}
}

@article{yazdani2019combined,
  title={A combined compromise solution (CoCoSo) method for multi-criteria decision-making problems},
  author={Yazdani, Morteza and Zarate, Pascale and Kazimieras Zavadskas, Edmundas and Turskis, Zenonas},
  journal={Management decision},
  volume={57},
  number={9},
  pages={2501--2519},
  year={2019},
  publisher={Emerald Publishing Limited}
}

@article{keshavarz2016new,
  title={A new combinative distance-based assessment (CODAS) method for multi-criteria decision-making},
  author={Keshavarz Ghorabaee, Mehdi and Zavadskas, Edmundas Kazimieras and Turskis, Zenonas and Antuchevi{\v{c}}ien{\.e}, Jurgita},
  year={2016},
  publisher={Academy of Economic Studies}
}

@article{saari1996copeland,
  title={The copeland method: I.: Relationships and the dictionary},
  author={Saari, Donald G and Merlin, Vincent R},
  journal={Economic theory},
  volume={8},
  pages={51--76},
  year={1996},
  publisher={Springer}
}

@article{zavadskas2007multi,
  title={Multi-attribute assessment of road design solutions by using the COPRAS method},
  author={Zavadskas, Edmundas Kazimieras and Kaklauskas, Artiras and Peldschus, Friedel and Turskis, Zenonas},
  journal={The Baltic journal of Road and Bridge engineering},
  volume={2},
  number={4},
  pages={195--203},
  year={2007}
}

@article{puvska2022evaluation,
  title={Evaluation and selection of healthcare waste incinerators using extended sustainability criteria and multi-criteria analysis methods},
  author={Pu{\v{s}}ka, Adis and Stevi{\'c}, {\v{Z}}eljko and Pamu{\v{c}}ar, Dragan},
  journal={Environment, Development and Sustainability},
  pages={1--31},
  year={2022},
  publisher={Springer}
}

@article{si2018dematel,
  title={DEMATEL technique: a systematic review of the state-of-the-art literature on methodologies and applications},
  author={Si, Sheng-Li and You, Xiao-Yue and Liu, Hu-Chen and Zhang, Ping},
  journal={Mathematical problems in Engineering},
  volume={2018},
  number={1},
  pages={3696457},
  year={2018},
  publisher={Wiley Online Library}
}

@article{keshavarz2015multi,
  title={Multi-criteria inventory classification using a new method of evaluation based on distance from average solution (EDAS)},
  author={Keshavarz Ghorabaee, Mehdi and Zavadskas, Edmundas Kazimieras and Olfat, Laya and Turskis, Zenonas},
  journal={Informatica},
  volume={26},
  number={3},
  pages={435--451},
  year={2015},
  publisher={Vilnius University Institute of Mathematics and Informatics}
}

@article{roy1968classement,
  title={Classement et choix en pr{\'e}sence de points de vue multiples},
  author={Roy, Bernard},
  journal={Revue fran{\c{c}}aise d'informatique et de recherche op{\'e}rationnelle},
  volume={2},
  number={8},
  pages={57--75},
  year={1968},
  publisher={EDP Sciences}
}

@article{shannon1948mathematical,
  title={A mathematical theory of communication},
  author={Shannon, Claude Elwood},
  journal={The Bell system technical journal},
  volume={27},
  number={3},
  pages={379--423},
  year={1948},
  publisher={Nokia Bell Labs}
}

@article{kuo2008use,
  title={The use of grey relational analysis in solving multiple attribute decision-making problems},
  author={Kuo, Yiyo and Yang, Taho and Huang, Guan-Wei},
  journal={Computers \& industrial engineering},
  volume={55},
  number={1},
  pages={80--93},
  year={2008},
  publisher={Elsevier}
}

@article{zavadskas2016integrated,
  title={Integrated determination of objective criteria weights in MCDM},
  author={Zavadskas, Edmundas Kazimieras and Podvezko, Valentinas},
  journal={International Journal of Information Technology \& Decision Making},
  volume={15},
  number={02},
  pages={267--283},
  year={2016},
  publisher={World Scientific}
}

@article{pamuvcar2015selection,
  title={The selection of transport and handling resources in logistics centers using Multi-Attributive Border Approximation area Comparison (MABAC)},
  author={Pamu{\v{c}}ar, Dragan and {\'C}irovi{\'c}, Goran},
  journal={Expert systems with applications},
  volume={42},
  number={6},
  pages={3016--3028},
  year={2015},
  publisher={Elsevier}
}

@article{marcelino2019development,
  title={Development of a multi criteria decision analysis model for pavement maintenance at the network level: Application of the MACBETH approach},
  author={Marcelino, Pedro and Antunes, Maria de Lurdes and Fortunato, Eduardo and Gomes, Marta Castilho},
  journal={Frontiers in Built Environment},
  volume={5},
  pages={6},
  year={2019},
  publisher={Frontiers Media SA}
}

@article{hadian2022multi,
  title={Multi attributive ideal-real comparative analysis (MAIRCA) method for evaluating flood susceptibility in a temperate Mediterranean climate},
  author={Hadian, Sanaz and Shahiri Tabarestani, Ehsan and Pham, Quoc Bao},
  journal={Hydrological Sciences Journal},
  volume={67},
  number={3},
  pages={401--418},
  year={2022},
  publisher={Taylor \& Francis}
}

@article{gligoric2022novel,
  title={Novel hybrid MPSI--MARA decision-making model for support system selection in an underground mine},
  author={Gligori{\'c}, Milo{\v{s}} and Gligori{\'c}, Zoran and Lutovac, Suzana and Negovanovi{\'c}, Milanka and Langovi{\'c}, Zlatko},
  journal={Systems},
  volume={10},
  number={6},
  pages={248},
  year={2022},
  publisher={MDPI}
}

@article{stevic2020sustainable,
  title={Sustainable supplier selection in healthcare industries using a new MCDM method: Measurement of alternatives and ranking according to COmpromise solution (MARCOS)},
  author={Stevi{\'c}, {\v{Z}}eljko and Pamu{\v{c}}ar, Dragan and Pu{\v{s}}ka, Adis and Chatterjee, Prasenjit},
  journal={Computers \& industrial engineering},
  volume={140},
  pages={106231},
  year={2020},
  publisher={Elsevier}
}

@article{brauers2006moora,
  title={The MOORA method and its application to privatization in a transition economy},
  author={Brauers, Willem Karel and Zavadskas, Edmundas Kazimieras},
  journal={Control and cybernetics},
  volume={35},
  number={2},
  pages={445--469},
  year={2006},
  publisher={Polska Akademia Nauk. Instytut Bada{\'n} Systemowych PAN}
}

@article{jagadish2014green,
  title={Green cutting fluid selection using MOOSRA method},
  author={Jagadish, Ray A and Ray, A},
  journal={International Journal of Research in Engineering and Technology},
  volume={3},
  number={3},
  pages={559--563},
  year={2014}
}

@article{madic2015selection,
  title={Selection of non-conventional machining processes using the OCRA method},
  author={Madi{\'c}, Milo{\v{s}} and Petkovi{\'c}, Du{\v{s}}an and Radovanovi{\'c}, Miroslav},
  journal={Serbian Journal of Management},
  volume={10},
  number={1},
  pages={61--73},
  year={2015}
}

@article{ataei2020ordinal,
  title={Ordinal priority approach (OPA) in multiple attribute decision-making},
  author={Ataei, Younes and Mahmoudi, Amin and Feylizadeh, Mohammad Reza and Li, Deng-Feng},
  journal={Applied Soft Computing},
  volume={86},
  pages={105893},
  year={2020},
  publisher={Elsevier}
}

@article{roubens1982preference,
  title={Preference relations on actions and criteria in multicriteria decision making},
  author={Roubens, Marc},
  journal={European Journal of Operational Research},
  volume={10},
  number={1},
  pages={51--55},
  year={1982},
  publisher={Elsevier}
}

@article{mufazzal2018new,
  title={A new multi-criterion decision making (MCDM) method based on proximity indexed value for minimizing rank reversals},
  author={Mufazzal, Sameera and Muzakkir, SM},
  journal={Computers \& Industrial Engineering},
  volume={119},
  pages={427--438},
  year={2018},
  publisher={Elsevier}
}

@article{maniya2010selection,
  title={A selection of material using a novel type decision-making method: Preference selection index method},
  author={Maniya, Kalpesh and Bhatt, Mangal Guido},
  journal={Materials \& Design},
  volume={31},
  number={4},
  pages={1785--1789},
  year={2010},
  publisher={Elsevier}
}

@article{vzivzovic2020eliminating,
  title={Eliminating rank reversal problem using a new multi-attribute model—the RAFSI method},
  author={{\v{Z}}i{\v{z}}ovi{\'c}, Mali{\v{s}}a and Pamu{\v{c}}ar, Dragan and Albijani{\'c}, Miloljub and Chatterjee, Prasenjit and Pribi{\'c}evi{\'c}, Ivan},
  journal={Mathematics},
  volume={8},
  number={6},
  pages={1015},
  year={2020},
  publisher={MDPI}
}

@article{hinloopen1990qualitative,
  title={Qualitative multiple criteria choice analysis: the dominant regime method},
  author={Hinloopen, Edwin and Nijkamp, Peter},
  journal={Quality and quantity},
  volume={24},
  pages={37--56},
  year={1990},
  publisher={Springer}
}

@article{cables2016rim,
  title={RIM-reference ideal method in multicriteria decision making},
  author={Cables, Elio and Lamata, Mar{\'\i}a T and Verdegay, Jose L},
  journal={Information Sciences},
  volume={337},
  pages={1--10},
  year={2016},
  publisher={Elsevier}
}

@article{madic2016application,
  title={Application of the ROV method for the selection of cutting fluids},
  author={Madi{\'c}, Milo{\v{s}} and Radovanovi{\'c}, Miroslav and Mani{\'c}, Miodrag},
  journal={Decision Science Letters},
  volume={5},
  number={2},
  pages={245--254},
  year={2016}
}

@article{panjaitan2019simple,
  title={Simple Additive Weighting (SAW) method in determining beneficiaries of foundation benefits},
  author={Panjaitan, Muhammad Iqbal},
  journal={Login},
  volume={13},
  number={1},
  pages={19--25},
  year={2019},
  publisher={SEAN Institute}
}

@article{keshavarz2018simultaneous,
  title={Simultaneous evaluation of criteria and alternatives (SECA) for multi-criteria decision-making},
  author={Keshavarz-Ghorabaee, Mehdi and Amiri, Maghsoud and Zavadskas, Edmundas Kazimieras and Turskis, Zenonas and Antucheviciene, Jurgita},
  journal={Informatica},
  volume={29},
  number={2},
  pages={265--280},
  year={2018},
  publisher={Vilnius University Institute of Mathematics and Informatics}
}

@article{olson1997decision,
  title={Decision aids for selection problems},
  author={Olson, David L},
  journal={Journal of the Operational Research Society},
  volume={48},
  number={5},
  pages={541--542},
  year={1997},
  publisher={Taylor \& Francis}
}

@inproceedings{dezert2020spotis,
  title={The SPOTIS rank reversal free method for multi-criteria decision-making support},
  author={Dezert, Jean and Tchamova, Albena and Han, Deqiang and Tacnet, Jean-Marc},
  booktitle={2020 IEEE 23rd International Conference on Information Fusion (FUSION)},
  pages={1--8},
  year={2020},
  organization={IEEE}
}

@article{khannoussi2022simple,
  title={Simple ranking method using reference profiles: incremental elicitation of the preference parameters},
  author={Khannoussi, Arwa and Olteanu, Alexandru-Liviu and Labreuche, Christophe and Meyer, Patrick},
  journal={4OR},
  pages={1--32},
  year={2022},
  publisher={Springer}
}

@article{gomes2009application,
  title={An application of the TODIM method to the multicriteria rental evaluation of residential properties},
  author={Gomes, Luiz Fl{\'a}vio Autran Monteiro and others},
  journal={European Journal of Operational Research},
  volume={193},
  number={1},
  pages={204--211},
  year={2009},
  publisher={Elsevier}
}

@article{hwang1981methods,
  title={Methods for multiple attribute decision making},
  author={Hwang, Ching-Lai and Yoon, Kwangsun and Hwang, Ching-Lai and Yoon, Kwangsun},
  journal={Multiple attribute decision making: methods and applications a state-of-the-art survey},
  pages={58--191},
  year={1981},
  publisher={Springer}
}

@article{opricovic2004compromise,
  title={Compromise solution by MCDM methods: A comparative analysis of VIKOR and TOPSIS},
  author={Opricovic, Serafim and Tzeng, Gwo-Hshiung},
  journal={European journal of operational research},
  volume={156},
  number={2},
  pages={445--455},
  year={2004},
  publisher={Elsevier}
}

@article{zavadskas2012optimization,
  title={Optimization of weighted aggregated sum product assessment},
  author={Zavadskas, Edmundas Kazimieras and Turskis, Zenonas and Antucheviciene, Jurgita and Zakarevicius, Algimantas},
  journal={Elektronika ir elektrotechnika},
  volume={122},
  number={6},
  pages={3--6},
  year={2012}
}

@article{san2012weighted,
  title={Weighted sum method and weighted product method},
  author={San Crist{\'o}bal Mateo, Jos{\'e} Ram{\'o}n and Mateo, Jos{\'e} Ram{\'o}n San Crist{\'o}bal},
  journal={Multi criteria analysis in the renewable energy industry},
  pages={19--22},
  year={2012},
  publisher={Springer}
}

@article{michnik2013weighted,
  title={Weighted Influence Non-linear Gauge System (WINGS)--An analysis method for the systems of interrelated components},
  author={Michnik, Jerzy},
  journal={European Journal of Operational Research},
  volume={228},
  number={3},
  pages={536--544},
  year={2013},
  publisher={Elsevier}
}

@article{stanujkic2021integrated,
  title={An integrated simple weighted sum product method—WISP},
  author={Stanujkic, Dragisa and Popovic, Gabrijela and Karabasevic, Darjan and Meidute-Kavaliauskiene, Ieva and Uluta{\c{s}}, Alptekin},
  journal={IEEE Transactions on Engineering Management},
  volume={70},
  number={5},
  pages={1933--1944},
  year={2021},
  publisher={IEEE}
}

@book{keeney1993decisions,
  title={Decisions with multiple objectives: Preferences and value tradeoffs},
  author={Keeney, Ralph L},
  year={1993},
  publisher={Cambridge university press}
}

@article{hodgett2014handling,
  title={Handling uncertain decisions in whole process design},
  author={Hodgett, Richard E and Martin, Elaine B and Montague, Gary and Talford, Mark},
  journal={Production Planning \& Control},
  volume={25},
  number={12},
  pages={1028--1038},
  year={2014},
  publisher={Taylor \& Francis}
}

@article{hodgett2019sure,
  title={SURE: A method for decision-making under uncertainty},
  author={Hodgett, Richard Edgar and Siraj, Sajid},
  journal={Expert Systems with Applications},
  volume={115},
  pages={684--694},
  year={2019},
  publisher={Elsevier}
}

@article{jacquet1982assessing,
  title={Assessing a set of additive utility functions for multicriteria decision-making, the UTA method},
  author={Jacquet-Lagreze, Eric and Siskos, Jean},
  journal={European journal of operational research},
  volume={10},
  number={2},
  pages={151--164},
  year={1982},
  publisher={Elsevier}
}

@article{salabun2015characteristic,
  title={The characteristic objects method: A new distance-based approach to multicriteria decision-making problems},
  author={Sa{\l}abun, Wojciech},
  journal={Journal of Multi-Criteria Decision Analysis},
  volume={22},
  number={1-2},
  pages={37--50},
  year={2015},
  publisher={Wiley Online Library}
}

\end{document}